\documentclass[twocolumn,twoside,slac_two]{revtex4}
\usepackage{graphicx}
\usepackage{fancyhdr}
\begin{document}

\title{Upsilon Decays into Scalar Dark Matter}

%

\author{Gagik K. Yeghiyan}
\affiliation{Department of Physics and Astronomy, Wayne State University, Detroit, MI 48201, USA}

\begin{abstract}
We examine decays of a spin-1 bottomonium into a pair of
light scalar Dark Matter (DM) particles, assuming that Dark
Matter is produced due to exchange of heavy
degrees of freedom.
We perform a model-independent
analysis
and derive formulae for the branching ratios of these decays. We
confront our calculation results with the experimental data.
We show that Dark Matter
production in $\Upsilon$ decays may lead to
constraints on parameters of the models containing a
light spin-0 DM particle.

\end{abstract}

\maketitle

\thispagestyle{fancy}


\section{Introduction}
\renewcommand{\theequation}{1.\arabic{equation}}
\setcounter{equation}{0}

We consider the possibility of using of
$\Upsilon$ meson decays with missing energy,
to test the models with a light spin-0 DM particle. Dark Matter search in
heavy meson (and in
particular $\Upsilon$ meson) decays may be complimentary to
such experiments as DAMA \cite{54,55,56},
CDMS \cite{57} and XENON \cite{58,59}, which rely on the measurement of
kinematic recoil of nuclei in DM interactions and lose (for cold DM particles)
sensitivity with decreasing mass of the WIMP, as the recoil energy becomes small.

So far, $\Upsilon$ meson decays into Dark Matter have been considered within the models,
where DM particles interaction with an ordinary matter is
mediated by some light degree of freedom \cite{48,37,36}. Apart from desire of having DM
annihilation enhancement (due to a light intermediate resonance) and thus having no
tension with the DM relic abundance condition \cite{34}-\cite{24},
it is also known that
$\Upsilon$ meson SM decay is predominantly due to strong interactions. Thus,
the
WIMP production branching ratio, in general, is greatly suppressed compared to
relevant weak B decays, and in particular
to $B \to K + invisible$ transition \cite{16,15}. In light of this,
it might seem natural to concentrate only on the models within
which Dark Matter production in $\Upsilon$ decays is enhanced due to
exchange of a light particle propagator.

Yet, our aim is to study $\Upsilon(1S)$ decay into a pair of spin-0 DM particles,
$\Upsilon(1S) \to \Phi \Phi^*$, and  $\Upsilon(3S)$ decay into a pair of
spin-0 DM particles and a photon, $\Upsilon(3S) \to \Phi \Phi^* \gamma$,
within the models
where light Dark Matter interaction with an ordinary matter is due to exchange of
{\it heavy} particles (with masses exceeding the bottomonium mass).
These models may
be free of tension related to satisfying
the DM relic abundance constraint as well \cite{34,16,68,69,70}.
Also, new experimental data
on $\Upsilon$ decays into invisible states have been reported by the BaBaR
collaboration \cite{66,21}. According to these data,
\begin{equation}
B(\Upsilon(1S) \to invisible) < 3 \times 10^{-4} \label{i2}
\end{equation}
and
\begin{equation}
B(\Upsilon(3S) \to \gamma + invisible) < (0.7 - 31) \times 10^{-6} \label{i3}
\end{equation}
where the interval in the r.h.s. of eq.~(\ref{i3}) is related to the choice of the final state missing mass.
These bounds are significantly stronger than those on invisible
$\Upsilon(1S)$ decays (with or without  a photon emission), reported previously
by Belle and CLEO \cite{64,65} and quoted by Particle Data Group \cite{1}. We show
that BaBaR
experimental data on $\Upsilon$ meson invisible decays
may constrain
the parameter space of light scalar Dark Matter models, even if
there is no Dark Matter production
enhancement due to light intermediate states.

We also illustrate that the study of
Dark Matter production in $\Upsilon$ decays allows us to test regions of
parameter space of
light spin-0 DM models that are inaccessible for B meson
decays with missing energy. It is also worth
mentioning that $\Upsilon$ decays are sensitive to
a wider range of WIMP mass than B decays. Thus, the study of
WIMP production in $\Upsilon$
decays is complementary to that
for B meson decays.

At the energy scales, associated with $\Upsilon$ decays,
heavy intermediate degrees of freedom may be
integrated out, thus leading to a low-energy effective theory of four-particle
interactions.
Our strategy would be deriving first model-independent formulae for the
$\Upsilon(1S) \to
\Phi \Phi^*$ and $\Upsilon(3S) \to \Phi \Phi^* \gamma$ branching ratios
within the low-energy effective theory.
Then, we confront our predictions with the experimental data, deriving
model-independent bounds in terms of the Wilson operator expansion
coefficients, as the parameters that carry the information on an
underlying New Physics model. Finally,
within a given model, using the matching conditions for the Wilson coefficients,
we translate these bounds into those on the relevant parameters of the considered
model.

The talk is based on the results presented in \cite{published} and
\cite{unpublished}.

\section{Model-Independent Analysis}
\renewcommand{\theequation}{2.\arabic{equation}}
\setcounter{equation}{0}

\begin{figure*}[t]
\centering
\includegraphics[width=80mm]{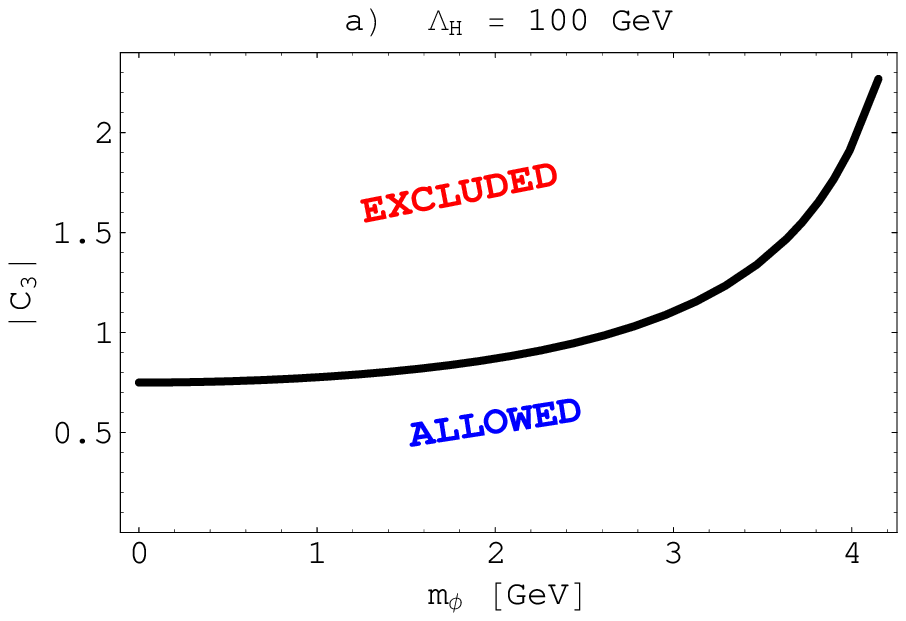}
\includegraphics[width=80mm]{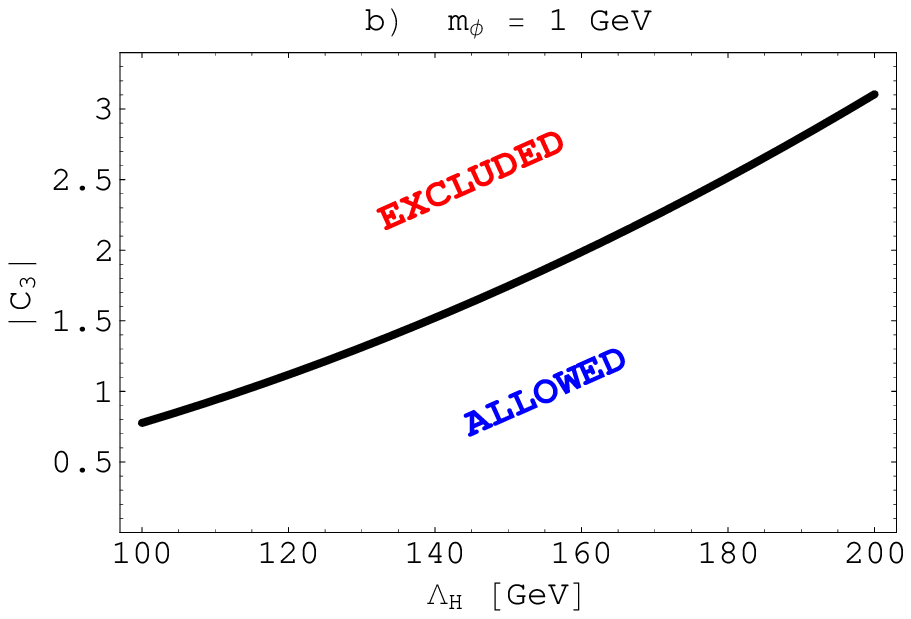}
\caption{Upper bound on $|C_3|$ a) as a function of $m_\Phi$, for $\Lambda_H = 100~GeV$,
b) as a function of $\Lambda_H$, for $m_\Phi = 1~GeV$.} \label{f7}
\end{figure*}

We treat $\Upsilon$ states - neglecting the sea quark and gluon
distributions - as
bound states of $b\bar{b}$ valence quark-antiquark pair that
annihilates - with or without emission of a photon - into a pair of
Dark Matter particles. To this approximation,
the relevant low-energy effective Hamiltonian may be written as
\begin{equation}
H_{eff} = \frac{2}{\Lambda_H^2} \sum_i{C_i \ O_i} \label{g1}
\end{equation}
where $\Lambda_H$ is the heavy mass  and
\hspace{-2.5cm}
\begin{eqnarray}
\nonumber
\hspace{-0.8cm} O_1 =  m_b \left(\bar{b} \ b \right) \left( \Phi^* \Phi \right), \
\ \ \ O_2 = i m_b \left(\bar{b} \gamma_5 b\right) \left(
\Phi^* \Phi \right), \ \  \\
\hspace{-0.8cm}
O_3 = \left(\bar{b} \gamma^\mu  b \right)
\left(\Phi^* i \partial^{^{^{\hspace{-0.22cm} \leftrightarrow}}}_\mu
\Phi \right), \ O_4 = \left(\bar{b} \gamma^\mu
\gamma_5 b \right) \left(\Phi^* i
\partial^{^{^{\hspace{-0.22cm} \leftrightarrow}}}_\mu \Phi \right)
\label{g2}
\end{eqnarray}
with $\partial^{^{^{\hspace{-0.22cm} \leftrightarrow}}} = 1/2
(\overrightarrow{\partial} - \overleftarrow{\partial})$.
It is worth noting that with the notations used in (\ref{g1})
and (\ref{g2}), all the operators $O_i$, i=1,..4, are Hermitean, thus all the
Wilson coefficients $C_i$ must be real.
If DM consists of
particles that are their own antiparticles, then only first two operators in eq.~(\ref{g2})
would contribute.

For DM field being a complex scalar state,
$\Upsilon(1S) \to \Phi \Phi^*$ branching ratio is given by the following expression:
\begin{eqnarray}
\nonumber
B(\Upsilon(1S) \to \Phi \Phi^*) =
\frac{\Gamma(\Upsilon(1S) \to \Phi \Phi^*)}{\Gamma_{\Upsilon(1S)}} = \\
 = \frac{C_3^2}{\Lambda_H^4} \frac{f_{\Upsilon(1S)}^2}{48 \pi \Gamma_{\Upsilon(1S)}}
\left[M_{\Upsilon(1S)}^2 - 4 m_\Phi^2 \right]^{3/2} \label{g5}
\end{eqnarray}
where $m_\Phi$ is the DM particle mass and
$\Gamma_{\Upsilon(1S)} $, $M_{\Upsilon(1S)}$, $f_{\Upsilon(1S)}$ are the
$\Upsilon(1S)$ total width, mass and decay constant respectively.
Only operator $O_3$ contributes to the decay rate.

For $\Phi$ being a self-conjugate spin-0 state, $\Phi = \Phi^*$,
contribution of $O_3$ vanishes as well, as noted above.
Thus, one has
\begin{equation}
B(\Upsilon(1S) \to \Phi \Phi) = 0 \label{g6}
\end{equation}
This result is related to the fact that
the final DM particle pair state must be a P-wave, which is impossible due to
the Bose-Einstein symmetry of identical spin-0 particles. In what follows,
$\Gamma(\Upsilon(1S) \to \Phi \Phi)$ must
also vanish in higher orders in $1/m_b$ operator product
expansion.

Thus, provided that DM pair production is the dominant invisible channel
(the neutrino background may be neglected \cite{published,64,66}),
the  signal for $\Upsilon(1S) \to  invisible$ decay would  imply that the light
spin-0 DM field has a complex nature. No evidence for the
$\Upsilon(1S) \to  invisible$ mode may lead to some constraints on
the parameters of the models with light complex scalar Dark Matter.

Yet, in order to derive such constraints, the experimental limit on  the
$\Upsilon(1S) \to invisible$ mode must be strong enough. Indeed, using
the numerical values of
$\Gamma_{\Upsilon(1S)} $, $M_{\Upsilon(1S)}$ and
$f_{\Upsilon(1S)}$ \cite{1}, \cite{19}, one may rewrite eq.~(\ref{g5}) as
\begin{eqnarray}
\nonumber
B(\Upsilon(1S) \to \Phi \Phi^*) \approx 5.3 \times 10^{-4} \ C_3^2 \times \\
\times \ \left(\frac{100GeV}{\Lambda_H} \right)^4 \
\left[1 - \frac{4 m_\Phi^2}{M_{\Upsilon(1S)}^2}
\right]^{3/2} \label{i10}
\end{eqnarray}
In what follows, the relevant experiments must be sensitive (at least)
to $B(\Upsilon(1S) \to invisible) \sim 10^{-4}$.

This sensitivity has been reached
by the BaBaR experiment \cite{66}, as it follows from the bound on
$B(\Upsilon(1S) \to invisible)$, given by eq.~(\ref{i2}). Substituting (\ref{i2})
into (\ref{i10}), one derives the following constraint on $|C_3|$ as a function of
$m_\Phi$ and $\Lambda_H$:
\begin{equation}
|C_3| < 0.75 \left(\frac{\Lambda_H}{100GeV}\right)^2
\left(1 - \frac{4 m_\Phi^2}{M_{\Upsilon(1S)}^2} \right)^{-3/4} \label{g25}
\end{equation}

The behavior of
the upper bound on $|C_3|$ with the DM particle mass and the heavy mass
is presented in Fig.'s~\ref{f7}a)~and~\ref{f7}b) respectively. For $m_\Phi < 3~GeV$ and
 $\Lambda_H \simeq 100~GeV$,
this bound may be translated into constraints on the relevant
couplings of models with a light complex spin-0 DM field. For $m_\Phi > 3~GeV$ or
for $\Lambda_H > 100~GeV$, further improvement of
the experimental sensitivity to $\Upsilon(1S) \to invisible$ transition
is necessary.

Note that even for $m_\Phi < 3~GeV$ and  $\Lambda_H \simeq 100~GeV$,
bound (\ref{g25}) still allows
$C_3$ to be of the order of unity. It
seems to be very unlikely to saturate
such a (rather weak) bound, if within a given New Physics model, $\Upsilon(1S) \to
\Phi \Phi^*$ transition is loop-induced. Thus, bound (\ref{g25})
on $|C_3|$ seems to be useful only within the models, where  $\Upsilon(1S) \to
\Phi \Phi^*$ decay occurs at tree level.

Further discussion of the scenarios with complex scalar Dark Matter goes
beyond the scope of this talk. Yet, in ref.~\cite{published} we use
bound (\ref{g25}) to derive constraints on the parameter space of
the models with mirror fermions, in particular on that of the MSSM with
gauge mediated SUSY breaking and the DM particle
in the hidden sector, while mirror fermions being connectors between the hidden
and the MSSM sectors. We derive for the first time bounds on the couplings of
b-quark interactions with its mirror counterpart, as functions of the WIMP mass
and the mirror fermion mass.

For the scenarios with the DM particle being its
own antiparticle, the decay
$\Upsilon(3S) \to \Phi \Phi \gamma$ is relevant.
As the neutrino background is negligible \cite{published},
$\Upsilon(3S) \to \Phi \Phi \gamma$ may be the dominant channel in
the $\Upsilon(3S) \to \gamma + invisible$ mode.

As it was mentioned above,
the strongest experimental constraint on
$B(\Upsilon \to \gamma + invisible)$ is that
derived by the BaBaR collaboration \cite{21} and
given by eq.~(\ref{i3}).
It may also be written as
\begin{equation}
B(\Upsilon(3S) \to \gamma + invisible ) < 3 \times 10^{-6} \label{i13}
\end{equation}
for the final state missing mass being less than 7~GeV, or approximately
$\sqrt{s} \lesssim M_{\Upsilon(3S)}/\sqrt{2}$. We have to note however that
this bound has been derived
assuming that $\Upsilon(3S) \to \gamma + invisible$ transition is mediated by
an intermediate resonant
Higgs state $A^0$ that exists in the MSSM extensions with an additional Higgs singlet
\cite{48,24,22}. For the models with non-resonant DM production considered here,
bound (\ref{i13}) may be used only to get {\it a preliminary estimate} of
possible constraints on parameters of light spin-0 self-conjugate DM
models\footnote{As for the earlier bounds on $B(\Upsilon \to \gamma + invisible)$ \cite{1},
the only existing constraint for the case of two-particle invisible states and the photon
having non-monochromatic energy,
$B(\Upsilon(1S) \to \gamma + X \overline{X}) < 10^{-3}$ \cite{65}, is too weak
to have any use. Due
to the factor $\alpha/(4 \pi) \approx 5.8 \times 10^{-4}$,
$B(\Upsilon \to \Phi \Phi \gamma)$ is always below this limit.}. More rigourously,
the experimental analysis performed in ref.~\cite{21} should be extended
to the cases, when the emitted photon energy is  non-monochromatic and is
in the range $0 < \omega < M_{\Upsilon(3S)}/2 -
2 m_\Phi^2/M_{\Upsilon(3S)}$. To our knowledge, this
work is in progress now\footnote{Author is grateful to Yu. Kolomensky for this
information.}.

\begin{figure}[t]
\centering
\includegraphics[width=80mm]{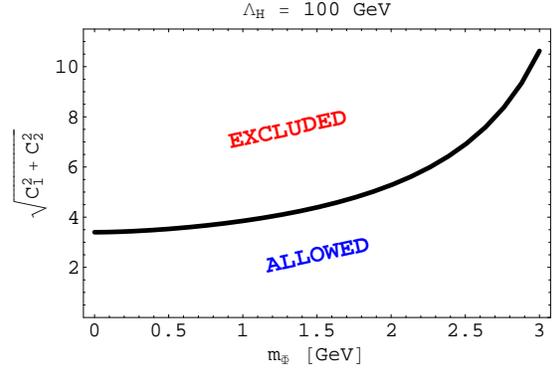}
\caption{Upper bound on $\sqrt{C_1^2 + C_2^2}$ as a function of
$m_\Phi$, for $\Lambda_H = 100~GeV$.} \label{f4}
\end{figure}

Within self-conjugate scalar DM scenarios,
the partially integrated branching ratio for $\Upsilon(3S) \to \Phi \Phi \gamma$
decay is given by
\begin{eqnarray}
\nonumber
&&\hspace{-0.7cm} B(\Upsilon(3S) \to \Phi \Phi \gamma)_{|s < s_{max}}  =
\frac{\left(C_{1}^2 + C_2^2\right)}{\Lambda_H^4}
\frac{\alpha}{4 \pi}\frac{f_{\Upsilon(3S)}^2}{54 \pi
\Gamma_{\Upsilon(3S)} M_{\Upsilon(3S)}} \times \\
\nonumber
&& \times \Biggl[ \Biggl(
2 M_{\Upsilon(3S)}^2 - s_{max} +
2 m_\Phi^2 \Biggr) \sqrt{s_{max} \left(s_{max} - 4 m_\phi^2\right)} - \\
&& \hspace{-0.7cm} - \ 8 m_\Phi^2 \left(M_{\Upsilon(3S)}^2 - m_\Phi^2 \right)
 \ \ln{\left(\frac{\sqrt{s_{max}} + \sqrt{s_{max}
- 4 m_\Phi^2}}{2 m_\Phi}\right)} \Biggr] \label{i8}
\end{eqnarray}
For $s_{max} = M_{\Upsilon(3S)}^2/2$, using the numerical values of the
$\Upsilon(3S)$ mass, total width and decay constant \cite{1,published},
one may rewrite eq.~(\ref{i8}) in the following form:
\begin{eqnarray}
\nonumber
&&\hspace{-1.5cm} B(\Upsilon(3S) \to \Phi \Phi \gamma)_{|s < M^2_{\Upsilon(3S)}/2} = \\
&&\hspace{-1cm} = 2.6 \times 10^{-7}
\left( C_1^2 + C_2^2 \right)
\left(\frac{100 GeV}{\Lambda_H} \right)^4 f(x_\Phi)
\label{i11}
\end{eqnarray}
where $x_\Phi = m_\Phi^2/M_{\Upsilon(3S)}^2$ and
\begin{eqnarray}
\nonumber
f(x_\Phi) = \left(1 + \frac{4}{3} x_\Phi \right) \sqrt{1 - 8 x_\Phi} \ - \\
- \ \frac{32}{3}
x_\Phi (1 - x_\Phi)
\ln{\left(\frac{1 + \sqrt{1 - 8 x_\Phi}}{2 \sqrt{2} \sqrt{x_\Phi}} \right)} \label{i12}
\end{eqnarray}

At first glance, it may seem that $\Upsilon(3S) \to \Phi \Phi \gamma$
branching ratio is
far out of reach of the BaBaR experimental sensitivity, for a reasonable choice of
$C_1$ and $C_2$.  Notice, however, that within
certain models with light spin-0 Dark Matter, the Wilson coefficients $C_1$ and/or $C_2$
may be enormously large, as they contain some enhancement factors, such as the ratio
$\Lambda_H/m_b \gg 1$ -  due to the mass term in the
numerator of a heavy fermion propagator, or the Higgs vev's ratio $\tan{\beta}$
(with the latter being, say, $\sim m_t/m_b \gg 1$) - due to DM particle pair production
via exchange of a heavy non-SM Higgs degree of freedom.

These enhancement factors can make $C_1$ and/or $C_2$ to be $\gtrsim 10$
and hence
$B(\Upsilon(3S) \to \Phi \Phi \gamma)$ to be $\sim 10^{-5} - 10^{-4}$, i.e.
significantly exceeding bound (\ref{i13}) on $B(\Upsilon \to \gamma + invisible)$.
Using bound (\ref{i13}) yields
\begin{equation}
\sqrt{C_1^2 + C_2^2} < 3.4 \left(\frac{\Lambda_H}{100 GeV}
\right)^2 f^{-1/2}(x_\Phi) \label{g27}
\end{equation}
The behavior of the bound on $\sqrt{C_1^2 + C_2^2}$ with the DM particle mass
is presented in Fig.~\ref{f4} for $\Lambda_H = 100~GeV$.

In the next section we consider the simplest possible model where
the Wilson coefficients are made enormously large due to an enhancement
factor. We transform the constraint on $\sqrt{C_1^2 + C_2^2}$ into
that on the relevant parameters of the model.

\section{Dark Matter Model with two Higgs Doublets (2HDM)}
\renewcommand{\theequation}{3.\arabic{equation}}
\setcounter{equation}{0}

\begin{figure}[t]
\centering
\includegraphics[width=80mm]{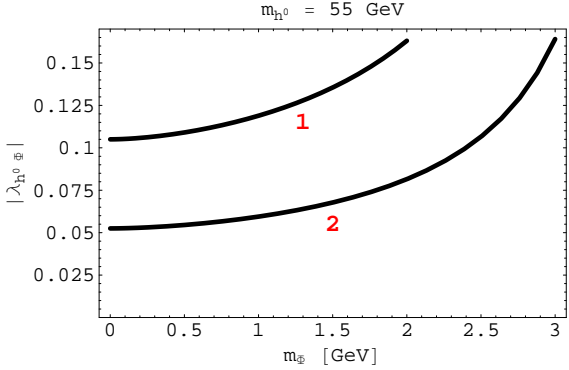}
\caption{Upper bound on $|\lambda_{h^0 \Phi}|$ as a function of $m_\Phi$, for $m_{h^0} = 55~GeV$
and $\tan{\beta} = 20$ (line 1), $\tan{\beta} = 40$ (line 2).} \label{f6}
\end{figure}

\begin{figure*}[t]
\centering
\includegraphics[width=80mm]{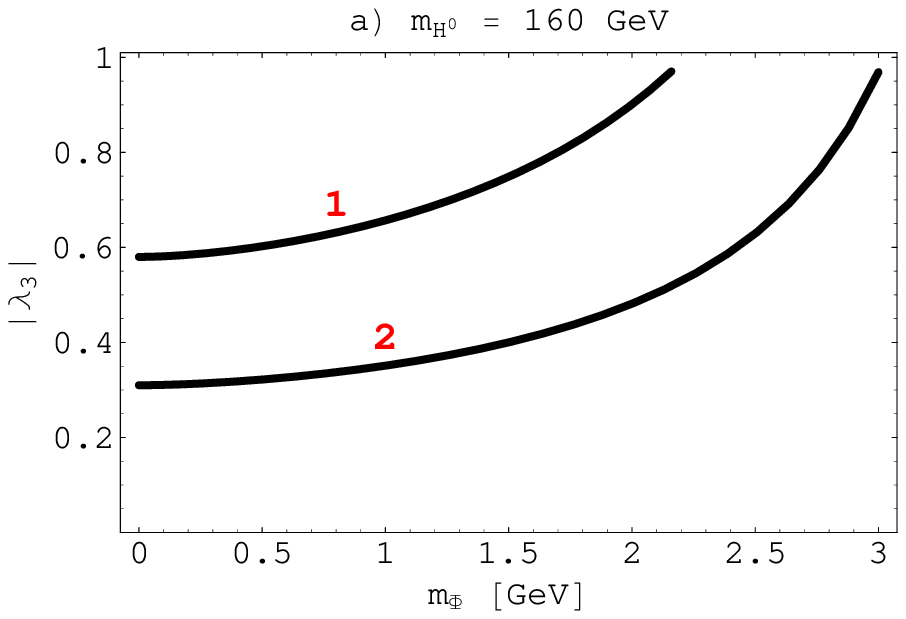}
\includegraphics[width=80mm]{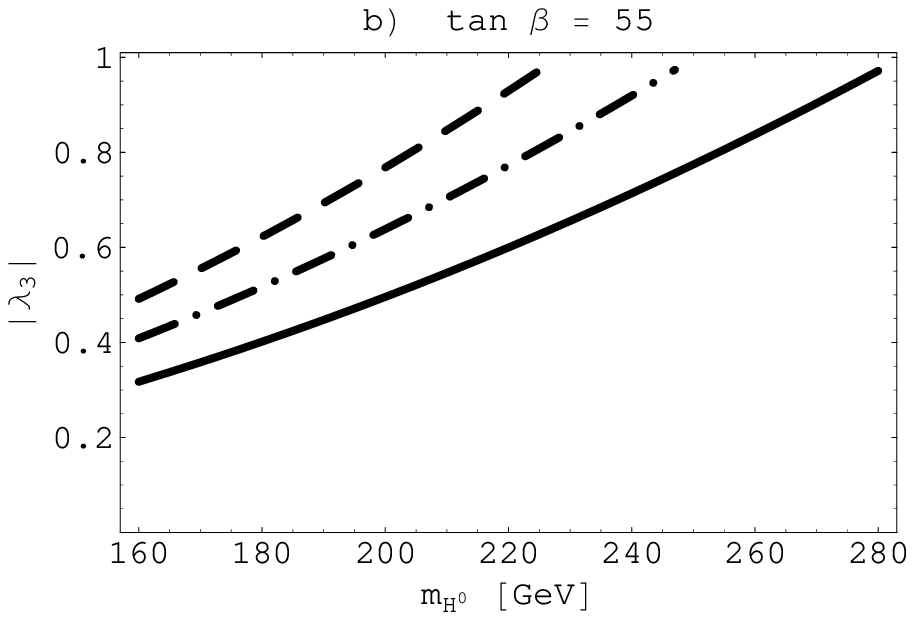}
\caption{Upper bound on $|\lambda_3|$, a) as a function of $m_\Phi$ for
$m_{H^0} = 160~GeV$ and $\tan{\beta} = 30$ (line 1), $\tan{\beta} = 55$
(line 2), b) as a function of $m_{H^0}$ for $\tan{\beta} = 55$ and
$m_\Phi = 100~MeV$ (solid line), $m_\Phi = 1.5~GeV$ (dashed-dotted line),
$m_\Phi = 2~GeV$ (dashed line).} \label{f5}
\end{figure*}

In this section we consider
the two-Higgs doublet model (2HDM) with
a gauge singlet real scalar DM particle.
The DM interaction part of Lagrangian, relevant for our analysis, may be written as
\cite{16}
\begin{eqnarray}
\nonumber
-{\cal L} = \frac{m_0^2}{2} \Phi^2 + \lambda_1 \Phi^2 |H_1|^2 +
\lambda_2 \Phi^2 |H_2|^2 + \\
+ \lambda_3 \Phi^2 \left(H_1 H_2 + h.c \right) \label{t1}
\end{eqnarray}
Following refs. \cite{16,9}, we consider type-II version of 2HDM,
where $H_1$ generates masses of down-type quarks and
charged leptons, whereas $H_2$ generates masses of up-type
quarks. We assume for the Higgs vev's ratio to be large,
i.e. $v_2/v_1 \equiv \tan{\beta} \gg 1$.

Three coupling constants entering Lagrangian~(\ref{t1}) are unknown parameters.
Constraints on the couplings
$\lambda_1$ and $\lambda_2$ may be derived from the study of
$B \to K + invisible $ transition. Those, in general, are strong enough:
combined with the ones coming from the DM relic abundance condition, they
rule out a wide range of the WIMP mass for the scenarios with $\lambda_1$ and/or
$\lambda_2$ dominant \cite{16}.

Yet, due to cancellation effects in the relevant diagrams, WIMP pair production
rate in B meson decays
is insensitive to the value of $\lambda_3$ \cite{16}.
The scenarios with $\lambda_3$
dominant, or at least non-negligible, are thus far unconstrained.
Study of DM production in $\Upsilon$ decays enables one
to constrain these scenarios,
inaccessible by B meson decays with missing energy.

The model contains two CP-even, one CP-odd
and two complex charged physical Higgs degrees of freedom.
To the leading order
in the perturbation theory,
$\Upsilon(3S) \to \Phi \Phi \gamma$ transition occurs at tree level by
exchange of a single Higgs boson. To this approximation,
only the CP-even Higgs states, $h^0$ and $H^0$, are relevant
for our analysis.

One of the CP-even Higgs bosons (not-necessarily the lightest one)
is Standard Model-like:
its phenomenology is similar to that of the SM Higgs boson and the experimental bound
on its mass is close to the SM limit\footnote{Also, if
the SM Higgs decays predominantly invisibly,
the SM lower experimental bound is distorted by a few GeV only \cite{29,30}.}
(see \cite{1} and references therein). Contribution of the diagrams with
the SM-like Higgs boson exchange to the $\Upsilon(3S) \to \Phi \Phi \gamma$
decay amplitude has no any enhancement factor - thus, we may further
disregard the Dark Matter interaction
with the SM-like CP-even Higgs.

The other CP-even Higgs boson is "New-Physics (NP) like": its phenomenology differs
drastically from that of
the SM Higgs boson \cite{25}. If Dark Matter is produced due to exchange of this
Higgs particle, $\Upsilon(3S) \to \Phi \Phi \gamma$
decay amplitude is enhanced
by $\tan{\beta}$ factor. As
discussed above, in this case the decay
branching ratio is within the reach of the BaBaR experimental
sensitivity.

If the NP-like Higgs is the lightest one, its mass
may be much
below the Standard Model experimental limit: according the existing experimental
data \cite{27,28}, $m_{h^0} > 55~GeV$ or $m_{h^0} < 1~GeV$ in the general type~II
2HDM. As it was mentioned above, the light Higgs scenario is beyond the scope of the
present paper, thus we assume here that $m_{h^0} > 55~GeV$.  Notice, however, that this
bound is derived, provided that no invisible Higgs decay mode exists. On the other hand,
if the NP-like Higgs invisible decay mode is dominant, it may escape
detection. No bound on $m_{h^0}$, to our best knowledge, exists in that case.

Within the considered model with light scalar Dark Matter,
analysis of the $\Upsilon \to \Phi \Phi \gamma$ mode may restrict
the scenarios with an invisibly
decaying lightest Higgs boson by putting severe constraints on the $h^0 \Phi \Phi$
interaction coupling,
$\lambda_{h^0 \Phi}$. For $\tan{\beta} \gg 1$ it is given by
\begin{equation}
\lambda_{h^0 \Phi} \approx \lambda_3 + \left(\lambda_1 +
\lambda_2 \right) \cos{\beta}  \label{l9}
\end{equation}
The last term in the r.h.s. of (\ref{l9}),
although being suppressed by a factor of
$\cos{\beta} \approx 1/\tan{\beta}$, must be retained because of
possible hierarchy in the values of $\lambda_3$ and $\lambda_1$ or $\lambda_2$.
Scenarios with such a hierarchy may be  of importance: using the matching
conditions for the Wilson coefficients,
\begin{equation}
C_1 = - \frac{\lambda_{h^0 \Phi}}{2} \tan{\beta}, \hspace{0.5cm}
C_2 = 0, \hspace{0.5cm} \Lambda_H = m_{h^0}, \label{t15}
\end{equation}
and bound (\ref{g27}) on $\sqrt{C_1^2 + C_2^2}$, one finds that
$\lambda_{h^0 \Phi}$ is constrained to be $O(1/\tan{\beta})$, if $h^0$ mass
approaches to its lower limit, $m_{h^0} = 55~GeV$. We illustrate this
in Fig.~\ref{f6} for $\tan{\beta} = 20$ and $\tan{\beta} = 40$.

For $\lambda_{h^0 \Phi}$ being so strongly constrained,
it seems to be very unlikely that $h^0$
would escape detection
and its mass be below 55 GeV.
More rigorously, however,
detailed re-analysis of the Higgs production and decay rates, including that of
$h^0 \to \Phi \Phi$,  should be performed.

One can show \cite{published} that when choosing $\tan{\beta}$ sufficiently large,
bound on $\lambda_{h^0 \Phi}$ may still be rigorous, if
the $h^0$ mass is heavier
than 55 GeV.
Thus, within the type~II 2HDM with a light spin-0 Dark Matter,
study of $\Upsilon \to \Phi \Phi \gamma$ decay channel may lead to severe constraints
on the lightest CP-even Higgs invisible decay coupling, if that Higgs is New-Physics like.

The remarkable feature of the model is that constraints on the parameter space
are derived even if the NP-like Higgs is the heaviest CP-even one. We illustrate
this choosing $m_{H^0} \geq 160 GeV$ - the existing theoretical upper bound
on the SM-like Higgs boson \cite{26} and exclusion of
the SM Higgs mass interval $(160 - 170)~GeV$ with 95\% C. L.
by the CDF and D0 data \cite{33} allow us to infer that
above 160 GeV, the CP-even Higgs boson is
presumably the heaviest one and NP-like. The matching conditions for the
Wilson coefficients are now
\begin{equation}
C_1 = \frac{- \lambda_3 \tan{\beta}}{2} \ , \hspace{0.5cm}
C_2 = 0 \ , \hspace{0.5cm} \Lambda_H = m_{H^0} \label{t22}
\end{equation}
Thus, bound (\ref{g27}) on $\sqrt{C_1^2 + C_2^2}$ may be transformed into that on
$|\lambda_3|$ as a function of the WIMP mass, $\tan{\beta}$ and the heaviest
CP-even Higgs mass.

As one can see from Fig.~\ref{f5}a),
for $m_{H^0} = 160~GeV$ and $\tan{\beta} = 30$, $\lambda_3$ is constrained to
be of order of the SM weak coupling or smaller ($|\lambda_3| \lesssim 0.65$), if
$m_\Phi \lesssim 1~GeV$. Also, for the same choice of the Higgs mass and
$\tan{\beta}$, $|\lambda_3|$ is to be less than one, if the WIMP mass is less than
2 GeV.
Bound on $|\lambda_3|$ is significantly more rigorous for higher values of
$\tan{\beta}$. For instance, if choosing
$\tan{\beta} = m_t(m_t)/m_b(m_t) \approx 55$, one gets
$|\lambda_3| \lesssim 0.35$
and $|\lambda_3| < 0.5$ for $m_\Phi \lesssim 1~GeV$ and $m_\Phi \simeq 2~GeV$
respectively.

The restrictions on $|\lambda_3|$ are essential also for higher values of
the heaviest CP-even Higgs mass: for $\tan{\beta} = 55$, they are still of
the interest
up to $m_{H^0} \simeq 280~GeV$, as one can see from Fig.~\ref{f5}b).

Thus, within the type~II 2HDM with a scalar Dark Matter, for large $\tan{\beta}$ scenario,
$\Upsilon$ meson decay into a Dark Matter particles pair and a photon, $\Upsilon \to
\Phi \Phi \gamma$, may be used to derive essential constraints on the parameters
of the model, which otherwise cannot be tested by B meson decays with invisible
outcoming particles.

\section{Conclusions and Summary}

Thus, spin-0 Dark Matter production in $\Upsilon$ meson decays has been investigated.
We restricted ourselves by consideration of the models where the decays occur due
to exchange of heavy non-resonant degrees of freedom.

We performed our calculations within low-energy effective theory, integrating out
heavy degrees of freedom. This way we derived model-independent formulae for
the considered branching ratios. We used these formulae to confront our
theoretical predictions with existing experimental data on invisible
$\Upsilon$ decays, both in a model-independent way and within particular
models. It has been shown that within the considered class of models, DM
production rate in $\Upsilon$ decays is within the reach of the present
experimental sensitivity. Thus, $\Upsilon$ meson decays into Dark Matter,
with or without a photon emission, may be used to constrain the models
with  a GeV or lighter spin-0 DM.

Experimental constraints on the $\Upsilon(3S) \to \gamma + invisible$ mode
are
derived assuming that Dark Matter is produced by exchange of a light
resonant scalar state. Within the scenarios with non-resonant DM
production considered here, these constraints may be used only to make preliminary
estimates of possible bounds on the parameters of the models. Yet,
those estimates show that these bounds may be rigorous enough; besides,
they are derived within the least restrained presently light scalar DM
scenarios.
Our goal is thus to encourage the experimental groups to analyze
the experimental data on $\Upsilon \to \gamma + invisible$ also
for the case of non-monochromatic photon emission and spin-0
invisible states.

So, from our analysis one may conclude that Dark Matter production in
$\Upsilon$ meson decays may serve as an interesting alternative to
commonly used DM search methods, capable of providing a valuable
information on DM particles, if those turn to have a mass of
the order of a few GeV or smaller.

\begin{acknowledgments}
Author is grateful to Yu. Kolomensky, A. Blechman,  B. McElrath and
J. Cao for valuable
comments and stimulating discussions. \\

This work has been supported by the grants
NSF~PHY-0547794 and DOE~DE-FGO2-96ER41005.

\end{acknowledgments}

\bigskip 

\end{document}